\documentclass{elsart}
\usepackage{graphicx}% Include figure files
\usepackage{dcolumn}% Align table columns on decimal point
\usepackage{amsmath,amsfonts}% popular packages from the American Mathematical Society
\usepackage{bm}% Bold Math package
\newcommand{\myint}[0]{\int \limits}
\newcommand{\intinfty}[0]{\int \limits_{-\infty}^{\infty}}

 \newcommand{\partiald}[2]{\frac{\partial #1}{\partial #2}}
\newcommand{\real}[1]{\mbox{Re}\left\{#1 \right\}}
\newcommand{\im}[1]{\mbox{Im}\left\{#1 \right\}}
\newcommand{\hypgeo}[2][p]{\: _#1\!\mathrm{F}_{\!#2}}

\begin{document}
\begin{frontmatter}
\title{Applications of the Flexural Impulse Response Functions in the Time Domain}

\author{Richard B\"ussow}
 \address{Institute of Fluid Mechanics and Engineering Acoustics, Berlin University of Technology, Einsteinufer 25, 10587 Berlin}
\ead{richard.buessow@tu-berlin.de}

\begin{abstract}
This work addresses the response functions for infinite beams and plates  with a
force excitation at the origin of the coordinate system and its application  to
time reversal and time frequency analysis. The response function for
Euler-Bernoulli beams is derived and for plates revisited. Interpretation
concerning energy conservation and mobility are given. An possible alternative
method to measure the coupling loss factors in SEA is sketched. \\
 The impulse response of a finite beam is measured in an experiment and compared
with that predicted theoretically. The experimental data before the first
reflection of the pulse showed good agreement with the theory. \\
The response function is used to simulate a time reversed pulse.  This numerical
simulation is verified with an experiment.  
\end{abstract}
                              %display desired
\end{frontmatter}

\section{Introduction}
This work addresses the response function for  an infinite beam and its
application to the time reversal in beams. The response function is an analytical function providing the responding velocity to ideally an impulse, but also to general excitation. \\
The time reversal technique \cite{fink92ieee,fink02phys} is originally invented in the context of ultrasonic sound and used mainly experimentally. The basic idea is first to record the impulse response of a source at many positions and then to playback this response. The result is a strong focused signal at the source position. The only prerequisite is reciprocity \cite{bojarski}. A theoretical study of the time reversal process in solids is in \cite{draeger} and experimentally in \cite{sutin}. The method is extended to other applications, e.g. damage detection \cite{gliozzi}. \\
Here the technique is applied to low frequency flexural waves in beams, for which the Euler - Bernoulli bending theory is valid. Whereas in \cite{draeger} a general theoretical formulation is found, here the process is studied theoretically by means of the response function of a homogeneous beam. The advantage of this restriction is that an analytical solution can be studied. This gives additional insight in the phenomena and opens new possibilities for this specialised case, since it is not necessary to record the impulse. \\
The application is restricted to beams, but nevertheless the corresponding function for an infinite plate is given. This shows that the concept can be easily extended to plates.\\
The time reversal is mainly an experimentell technique. A theoretical approach that is similar to the presented can be found in \cite{folkowgreen,folkowinverse} for the Timoshenko beam. In the study \cite{folkowgreen} the Green's function is obtained numerically whereas here an analytical function is studied. \\
The response function itself is also useful for the analysis of the time - frequency distribution of a flexural impulse response. This application is the subject of a related publication \cite{bendingwavtheo} and is shortly introduced in section \ref{timefreqsec}. \\ 
The function itself was first derived for teaching purpose, since it is a instructive example for dispersive wave propagation. In this context the functions are discussed with respect to energy considerations and the mobility in section \ref{energymobisec}. This provides interesting insight since well known phenomena are highlighted from a different point of view. \\
The expectation is that given its basic character the response function should
exist. An  intensive search in mainly old scientific documents revealed the
Green's functions  for an initial deflection and velocity that is stated in the
next section.  From a phenomenological point of view there is maybe nothing to
add  since these functions are very similar. For a practical application there
is a fundamental difference since for the later derived impulse response it is
not necessary to know the initial deflection and velocity of the beam that
result from the impulse. 
\subsection{Green's function}
Green's functions for beams and plates were first derived  by Boussinesq
\cite{boussinesq}. The governing equation for the deflection $\xi(x,t)$ of a beam is
\begin{equation}
B \partiald{^4 \xi}{x^4} + m' \partiald{^2 \xi}{t^2} = 0,
\label{biegedgl}
\end{equation}
where $x$ is the distance from the source, $\zeta=\sqrt{B/m'}$, where $B$ is the flexural rigidity and $m'$ mass per unit length.\\
The Green's function is obtained by a Laplace transform of the differential equation (\ref{biegedgl}) and its initial and boundary conditions. The Laplace transform and its application to transient vibration is well explained for example by Thomson \cite{thomson}. With the initial conditions
\begin{equation}
\xi(0,t) = V(t);  \: \xi(x,0) = 0; \: \dot{\xi}(x,0)=0 
\end{equation}
for the deflection $\xi(x,t)$, it holds that
\begin{equation}
\begin{array}{l}
\xi(x,t) =  \\
\frac{x}{4 \pi} \sqrt{\frac{2 \pi}{\zeta}} \myint_{0}^{t} \frac{V(\tau)}{(t-\tau)^{3/2}} \left( \sin \frac{x^2 }{4 \zeta ( t - \tau )} + \cos \frac{x^2}{4 \zeta (t - \tau) } \right) \: d \tau.
\end{array}
\label{xiboussi}
\end{equation}
 This function and its derivation can be found in several textbooks e.g.
 \cite{love,nowacki}. For applying  the Laplace transform it is necessary to
 know the initial conditions.  It is not obvious how to incorporate the
 excitation. For a plate, Boussinesq assumed  that the initial deflection from
 a concentrating force  acting at the origin of the coordinate system is
 $\xi(r,0) = \frac{\delta(r)}{2 \pi r}$.  This does not apply for the infinite
 beam as is shown  later, equations see (\ref{vorigin}) and (\ref{xibeam}).   

\section{Response Functions}
In this section the response functions for beams and plates are derived  and
discussed. In the following equation (\ref{biegedgl}) is solved for the 
velocity $v$ and not anymore for the deflection. The simple solution
(\ref{vdirac})  that is derived in the following is only obtained if the
velocity is chosen  as the unknown variable. The approach is identical to that
used in \cite{heckl}.  The difference to the example that is given in
\cite{heckl} is that the mobility  is not altered by a sticking mass that impacts the beam. 
 
\subsection{Semi-Infinite and Infinite Beams}
Consider a theoretical setup of a semi-infinite $(x \in [0, \infty])$ or infinite beam, which is excited at $x=0$ with an arbitrarily short force pulse $F_a$. The pulse is modelled by the Dirac $\delta$-function $F_a(t) = F_0 \: \delta(t)$ whose Fourier transform is given by
\begin{equation}
\hat{F}_a(\omega) = \myint_{-\infty}^T F_a(t) e^{j \omega t} \: d t = F_0 H(T) 
\end{equation}
where $H(T)$ is the Heaviside-Function. To find the response function to the impulse one may proceed with the boundary conditions for this elementary problem. The equations for the angular velocity $w$, bending moment $M$, shear force $F_y$ and velocity $v$ of a beam which can be modelled with the Euler beam theory, equation (\ref{biegedgl}) are 
\begin{equation}
\hat{w} = \partiald{\hat{v}}{x}, ~ \hat{M} = - \frac{B}{j\omega} \partiald{\hat{w}}{x}, ~ \hat{F}_y = - \partiald{\hat{M}}{x}, ~ j \omega m' \hat{v} = - \partiald{\hat{F}_y}{x}.
\end{equation}
Further is the bending wave number $k_b= \omega/c_b $, the bending wave velocity $c_b = \sqrt[4]{\omega^2 B/m'}$, the flexural rigidity  $B=E I_y$, $E$ the elastic or Young's modulus, $I_y$ the geometrical moment of inertia. 
\subsubsection{Boundary Conditions}
For a semi-infinite beam the waves propagate away from the excitation point which leads with $v=\real{\hat{v} e^{j \omega t}}$ to
\begin{equation}
\hat{v} = \hat{v}_+ e^{-j k_b x} + \hat{v}_{+j} e^{-k_b x}.  
\end{equation}
Herein the term $\hat{v}_+$ is the amplitude of the far-field wave and $\hat{v}_{+j}$  that of the near-field wave. At the free end the bending moment and shear force must vanish $F(x=0)=F_a$ and $M(x=0)=0$. It follows with $1+j = \sqrt{2j}$ that 
\begin{equation}
\hat{v}(x,\omega) = \frac{\hat{2 F_a}(\omega)}{ \sqrt{2 j} \, m' c_b }\left( e^{- j k_b x} + e^{- k_b x} \right). 
\label{vom}
\end{equation}
In case of an infinite beam symmetry implies that
\begin{equation}
\hat{v} = \hat{v}_+ e^{-jk |x|} + \hat{v}_{+j} e^{-k |x|}.  
\end{equation}
At $x=0$ the angular velocity is $w(x=0)=0$ and $F_a/2=F_y(x=0)$, therefore
\begin{equation}
\hat{v}(x,\omega) = \frac{\hat{F_a}(\omega)}{ 2 m' c_b \sqrt{2j} }\left( e^{- j k_b |x|} +  e^{- k_b |x|} \right).
\label{vuend}
\end{equation}
This is the same result as in equation (\ref{vom}), except for the factor $1/4$. 

\subsubsection{Inverse Fourier Transform} 
To obtain the response function $v(x,t)$ one may use the real part of the inverse Fourier transform of equation (\ref{vom})
\begin{equation}
v(x,t) = \real{\frac{1}{2 \pi} \intinfty \hat{v}(x, \omega) e^{i \omega t} \: d\omega},
\end{equation}
which with $\omega/\sqrt[4]{\omega^2}= \frac{\omega}{|\omega|} \sqrt{|\omega|}$ ensures that the waves always propagate away from the excitation point. Upon substitution the expression reads
\begin{equation}
\begin{array}{l}
v(x,t) = \frac{F_0 H(t)}{\pi m'\sqrt{\zeta}} \times \\
\real{\intinfty \frac{e^{-j \frac{\omega}{|\omega| \sqrt{\zeta}}  \sqrt{|\omega|}  x } + e^{-\frac{\omega}{|\omega| \sqrt{\zeta}}  \sqrt{|\omega|} x} }{\sqrt{2 j |\omega| } } e^{j \omega t} \: d\omega}
\end{array}
\label{vt}
\end{equation}
where $\zeta = \sqrt{B/m'}$. This integral is solved with Mathematica's {\tt Integrate[]} command with use of the hyper-geometric function $_pF_q$ for $t>0$. The hyper-geometric functions were found by Pochhammer\cite{pochhyp}. Surprisingly he also solved the problem of bending waves in a circular beam without assuming that the radial movements can be neglected \cite{timoshenkohist,pochbiege}.  \\
 He gives also an exact expression for the bending wave velocity in this general case \cite{pochvelo}. An overview of Pochhammer's work and its application towards transient response can be found in Miklowitz \cite{miklowitz}. \\
In the case of $t<0$ the integral vanishes due to the Heaviside-function. One thus finds that 
\begin{equation}
\begin{array}{l}
\intinfty \frac{e^{-j \frac{\omega}{|\omega| \sqrt{\zeta} } \sqrt{|\omega|} x } + e^{- \frac{\omega}{|\omega| \sqrt{\zeta}} \sqrt{|\omega|} x } }{\sqrt{2 j \omega} } e^{j \omega t} \: d\omega = \\ 
\sqrt{ \frac{ 2 \pi}{t}} \cos \frac{x^2}{4 \zeta t}-\\
 j \left( \frac{j 2 x}{\sqrt{\zeta} t} \: \hypgeo[1]{2} (1, \frac{3}{4} \: \frac{5}{4}, -\frac{x^4}{64 \zeta^2 t^2}) -  \sqrt{ \frac{ 2 \pi}{t}}  \cos \frac{ x^2}{4 \zeta t} \right).
\end{array}
\label{nahundfern}
\end{equation}
The pulse is an even function, so just the real part remains 
\begin{equation}
\fbox{$ \displaystyle v(x,t) = \frac{F_0 H(t)}{ m'}  \sqrt{ \frac{2 }{\pi \zeta t}} \cos \frac{ x^2}{4 \zeta t}$} \: .
\label{vdirac}
\end{equation}
It should be mentioned that to produce this result it is vital to use the near- and far-field terms in equation (\ref{vt}). For the origin at $x=0$, it follows that
\begin{equation}
v(x=0,t) = \frac{F_0 H(t)}{ m'}  \sqrt{ \frac{2 }{\pi \zeta t}}.
\label{vorigin}
\end{equation}
Since $\lim_{x,t \rightarrow 0} \frac{x^2}{4 \zeta t} = 0$, the  cosine is unity
and it follows for the velocity that $\lim_{x,t \rightarrow 0} v(x,t)
\rightarrow \infty$. This  relation can be found approximately in \cite{heckl}
for the example of a sticking mass. This example is maybe an indication that at
least in Cremer's time  the function (\ref{vdirac}) was not known, since a
sticking mass is rather  unusual and the force excitation has the advantage
that it can be  studied with an analytical function. \\
The primitive function of equation (\ref{vdirac}) gives the deflection
\begin{equation}
\xi(x,t) = \frac{2 F_0 H(t)}{ m'} \left(\sqrt{ \frac{2 t}{\pi \zeta}} \cos \frac{x^2}{4 \zeta t} + \frac{x}{\zeta}  S \left(\frac{x}{2 \pi c t} \right) \right),
\label{xibeam}
\end{equation}
where $S(x) = \int_0^x \sin(\pi t^2 /2) \: dt$ is the Fresnel-Sine-Integral. 
 
\paragraph{General excitation}
The Fourier transform $\hat{\Psi}(\omega)$ of a function $\psi(t)= \gamma(t) \times \phi(t)$ is
\begin{equation}
\hat{\Psi}(\omega) = \frac{1}{2 \pi} \hat{\Gamma}(\omega) \ast \hat{\Phi}(\omega) 
\end{equation}
where $\ast$ denotes the convolution. In case of an inverted Fourier transform there is no factor $1/(2 \pi)$ and it follows that
\begin{equation}
v(x,t) = F_a(t) \ast \frac{1}{m'}  \sqrt{ \frac{2} {\pi \zeta t}} \cos \frac{x^2}{4 \zeta t}.
\label{vfaltung}
\end{equation}
For an arbitrary force $F_a(t)$, $\im{\hat{F}_a(\omega)}=0$ does not hold. But $F_a(t)$ is always real and $v(x,t)$ is real such that only the real part of equation (\ref{nahundfern}) needs to be taken into account. It is a straightforward test to use the Dirac $\delta$-function as the force and to obtain
(\ref{vdirac}). 

\subsection{Infinite Plate}
The derivation of the response function for the infinite plate is given by Crighton \cite{crighton,meirovitch}. Nevertheless this problem is revisited, since the equation is used in the following. The starting point is the propagation function of an infinite plate
\begin{equation}
\hat{v}(\omega,r) = \frac{\hat{F}(\omega)}{8 \sqrt{ B' m''} } \left( H_0^{(2)}(k_b r) - H_0^{(2)}(-j k_b r) \right),
\label{plattevom}
\end{equation}
with the Hankel function of the second kind $H_0^{(2)}(x) = J_0(x) - j Y_0(x)$, the flexural rigidity of a plate $B'= \frac{E h^3}{12 (1-\nu)}$, $\nu$ the Poisson ratio and $m''$ the mass per unit area. The inverse Fourier transform can be simplified with the following identies of the Hankel function. For $x>0$ are
\begin{equation}
\begin{array}{ll}
H_0^{(2)}(-jx) = - j \real{Y_0(-j x)}, & J_0(x) = J_0(-x), \\
\im{Y_0(-x)} = 2 J_0(-x), & \im{J_0} = 0,   \\  
\real{Y_0(j x)} = \real{Y_0(-j x)},  & \im{Y_0(x)} = 0, \\
 \real{Y_0(x)} = \real{Y_0(-x)} &.  
\end{array}
\label{symmhankel}
\end{equation}
With (\ref{symmhankel}) the real part of the inverse Fourier transform of equation (\ref{plattevom}) reduces to 
\begin{equation}
v(r,t) = \frac{\hat{F}_0}{4 \pi \sqrt{ B' m''}} 
\myint_0^\infty J_0(\sqrt{\omega/\zeta } r   ) \cos(\omega t) \: d\omega.
\end{equation}
Hence, the response function for the infinite plate is
\begin{equation}
v(r,t) = \frac{\hat{F}_0}{4 \pi t \sqrt{ B' m''} } \sin \frac{r^2}{4 \zeta t}.
\label{schnelleplatte}
\end{equation}

\section{Dispersion factor and number}
An important difference between the bending wave and, for example, a longitudinal wave is that the group velocity is frequency dependent and so the response to an impulse is spreading into the different spectral fractions of the pulse. The term 
\begin{equation}
d_i=\frac{x^2}{4 \zeta}.
\label{di}
\end{equation}
is the factor that controls this spreading and is called dispersion factor. \\
Whereas the dispersion factor is a time value the nondimensional term  
\begin{equation}
Di=\frac{x^2 f_{max}}{4 \zeta},
\label{Di}
\end{equation}
is called dispersion number. In an experiment that should reveal the influence of the dispersion it is necessary to choose a structural sample with a high dispersion number. The maximum frequency is impotant since the higher the frequency the more the impulse will spread. \\
A high dispersion is the reason for choosing a thin plate and a slender beam.

\section{Time - Frequency Analysis}
\label{timefreqsec}
The time-frequency analysis of dispersive waves is widely studied and can be found in several publications. This section is only meant to give an overview of applications of the response function in this area. \\
One may insert $x=c_g t$ in equation (\ref{vdirac}), to obtain
\begin{equation}
v(\omega, t) = \frac{F_0 H(T)}{m'} \sqrt{\frac{ 2 }{ \pi \zeta t}} \cos \omega t,
\label{veloomt}
\end{equation} 
where the bending wave group velocity is $c_g= 2 c_b = 2 \sqrt{\omega} \zeta$. This shows the known fact that the frequency content of the bending wave is travelling with its particular group velocity. It is the theoretical basis for the time-frequency analysis of dispersive waves. For example Kishimoto et. al. \cite{kishimoto} show the validity of their approach with two neighbouring frequency components.\\
Often the wavelet transform \cite{kishimoto,gaulident} is used to extract the frequency dependent arrival time $t_a=c_g/x$.  \\
The time and frequency dependent energy is obtained from the signal by means of the wavelet transform. The proceduce to extract the arrival time is to find  the time value with the maximum energy at a fixed frequency. It is not possible to find the frequency value with the maximum energy at a fixed time, since this is e.g. affected by the non-constant amplitude $F_0$ of a real application. \\
An extention of the described method of time- frequency analysis is done by Vries et. al.\cite{vries}. The basic idea is to remove the dispersion by applying the response function. The response function itself is obtained by a numerical inverse Fourier transform. \\
Another approach is to use the analytical response function (\ref{vdirac}) as a wavelet mother function. The dispersion factor is then obtained by the scaling factor of the wavelet with the highest energy. This approach can be found in \cite{bendingwavtheo} and is applied to the signals obtained in the experiments that are presented in the following. \\
That the approach above is valied can also be shown with the instaneous frequency $\omega(t)$ of an almost periodic functions, see Bochner \cite{bochnerfp}. Bochner \cite{bochnerdgl} also showed that these functions are solution to the wave equation and also for equation (\ref{biegedgl}) that is used here. It holds that  
\begin{equation}
\cos \varphi(t) \rightarrow \omega(t) = \varphi'(t) = \frac{x^2}{4 \zeta t^2}.
\label{bochnerfreq}
\end{equation}

\section{Energy Conservation and Mobility}
\label{energymobisec}
The term $\sqrt{\frac{1}{t}}$ in equation (\ref{vdirac}) stems from the general energy conservation scaling of a function $f(x)$, which is
\begin{equation}
f_a(x)=\frac{1}{\sqrt{|a|}} f\left( \frac{x}{a} \right).
\end{equation}
One may interpret equation (\ref{veloomt}) so, that the pulse while traveling along the beam is scaled by the travel time such that the energy of the pulse is conserved. In the case of a plate the corresponding term is $\frac{1}{t}$, which is due to the fact that the pulse propagates cylindrically and not as a plane wave. The radius of the cylinder follows a $r \sim \sqrt{t}$ dependence. 
The complex mobility is  defined by
\begin{equation}
\hat{Y}  = \hat{v}/\hat{F}_a. 
\label{mobi}
\end{equation}
If the velocity and the force are not at the same position it is called transfer mobility.\\
For a semi-infinite beam the transfer mobility to a position in the far-field ($k_b x \gg 1$) is obtained as 
\begin{equation}
\hat{Y}(\omega) = \frac{2}{\sqrt{\omega \zeta} m' (1 + j)} e^{- j k_b x}. 
\end{equation}
The decrease in magnitude over frequency is given by
\begin{equation}
|Y(\omega)| = \frac{1}{m'} \sqrt{\frac{2}{\zeta}}  \sqrt{\frac{1}{\omega}}.
\label{admittanz} 
\end{equation}
That this corresponds to the solution in the time domain is discussed in the
following.  Consider the envelope of equation (\ref{vdirac}) to be
\begin{equation}
v_{env} = \frac{F_0 H(t)}{ m'}  \sqrt{ \frac{2 }{\pi \zeta t}}
\label{envelope}
\end{equation}
The Fourier transform of equation (\ref{envelope}) produces with
$FT\{\sqrt{1/t}\} = \sqrt{\pi/\omega}$ exactly  the same mobility as equation
(\ref{admittanz}). This indicates  that the remaining cosine-term in equation
(\ref{vdirac}) is not affecting  the magnitude of the mobility in the
far-field. From equation (\ref{freqsin})  follows that the far-field condition
($k_b x \gg 1$) for equation  (\ref{vdirac}) is $d_i / t \gg 1$. With this
prerequisite it follows  that the envelope (\ref{envelope}) defines the
amplitude of a frequency  component given by the remaining cosine-term, like in
equation (\ref{veloomt}).  For low frequencies $d_i/t \ll 1$ it is not possible
to analyse the two terms  independently, since the prerequisite for almost
periodic functions is  violated \cite{bochnerfp}. Nevertheless the Fourier
transform of equation  (\ref{vdirac}) results obviously in equation (\ref{vom}). \\
With this interpretation one can define a mobility in the time domain 
\begin{equation}
Y_t = v(t) / \hat{F}_a \left( \omega = \frac{x^2}{4 \zeta t^2} \right).
\label{mobilitytd}
\end{equation}
This definition is followed in the experimental section \ref{expbp}. It also leads to a possible application that is discussed in the following. 
% \subsection{plate}
% The relation corresponding to equation (\ref{admittanz}) for plates in the far-field ($k_b r \gg 1$) is given by
% \begin{equation}
% |Y(\omega)| = \frac{1}{8 m''} \sqrt{\frac{2}{\zeta \pi r}} \sqrt[4]{ \frac{1}{\omega}} .
% \label{admittanzplatte}
% \end{equation}

% From the energy conservation and the mobility one may speculate that the response function for a general dispersion relation $c_g = \zeta_g \omega^{1/n}$ for a one dimensional propagation has the structure of 
% \begin{equation}
% v(x,t) = F_0 \real{ \hat{A} \sqrt{t^{1-n}} e^{i x^{n}  t^{1-n}  \zeta_g^{-n}}},
% \end{equation}
% where $\hat{A}$ depends on the mobility. The formular affirms the energy conservation and that the wave travels with the group velocity.   
\subsection{Application to SEA}
It is shown that the mobility is trivially related to the response function. This opens a possible application in the context of the Statistical Energy Analysis SEA \cite{craik}. To measure the coupling loss factors used in a SEA model a standard method is to excite each subsystem in turn and measure the response of all the subsystems for each source \cite{craik,hodges}. \\
The elaborated interpretation opens an alternative method: excite a subsystem A with an impulse and measure the transient impulse reponse of the neighbouring subsystem B. The transmission coefficient can be easily obtained with the energy of the impulse and its response. The only prerequisite is that the direct transmission can be clearly distinguished from the early reflections. That this is in principle possible is shown in the experiments in section \ref{expbp}, where the response function for an infinite beam is verified with a finite beam. 

\section{Experimental results}
\label{expbp}
Measurements are carried out on a thin acrylic plate and a slender aluminum beam for different distances and configurations. The results show the same tendency. For the sake of brevity just one typical measurement of the  beam  and the plate is presented and discussed. 

\subsection{Beam}
\label{expbeam}
\begin{figure}
\includegraphics[width=0.9 \textwidth]{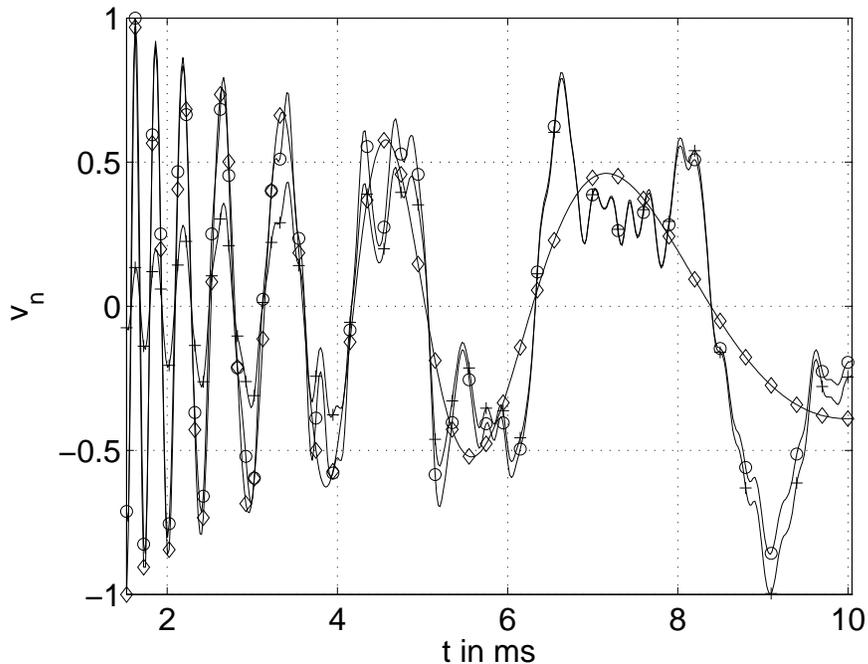}
\caption{Impulse response of a beam, normalised velocity $v(x=1.5m,t)/|v(x=1.5m,t)|_{max}$: measured (plus), theoretical (diamond) with equation (\ref{vdirac}), measured and corrected with the power spectrum of the impulse (circle)}
\label{veloclam3}
\end{figure}

The dimensions of the beam are a diameter of $d=6mm$,  a length of $l=3m$ and a
density of $\rho=2830 kg/m^3$. The material parameters for aluminium are tested
with resonance frequencies of the beam and a elastic modulus $E=68,0 GPa$ is obtained. \\
The beam is clamped at both ends. The velocity is measured with a
laser-vibrometer with a  sampling frequency of $50kHz$ placed at $0.5m$ from
the end. The beam is  excited by means of an impacting hammer equipped with a
force transducer at a  distance of $x=1.50m$ from the point of measurement. \\
Since the real beam is not infinite only the first passage of the pulse is
considered. The beam velocity  is plotted in figure \ref{veloclam3}. The time
axis is started at the maximum  of the force signal minus the delay of the
laser-vibrometer of $1.2 ms$. One may recognise that already after $3 ms$ the
reflections  from the clamped end are visible.\\
The theoretical curve is calculated by means of equation (\ref{vdirac}) with a
value of $d_i=0.079$.  This corresponds quite well with the value obtained with
equation (\ref{di}) of  $d_i=0.076$. The actual value is extracted with a
method that is discussed  in a subsequent publication. The frequency range of
the theoretical  curve is $f_{min}=124 Hz$ to $f_{max}=5.4 kHz$. The measured
curve is corrected  by means of the frequency distribution of the theoretical
curve that is  obtained from equation (\ref{freqsin}) and the power spectrum of
the  measured force impulse, like defined in equation (\ref{mobilitytd}). \\
%This correction can be interpreted as a mobility in the time domain. This method is not found in the literature and shows an application of the interpretation given in the previous section \ref{energymobisec}. \\
The curves are normalised with their maximum value, since in this context the 
distribution of amplitude and frequency over time is of interest, but not the absolute value.

\subsection{Plate}
\begin{figure}
\includegraphics[width=0.9 \textwidth]{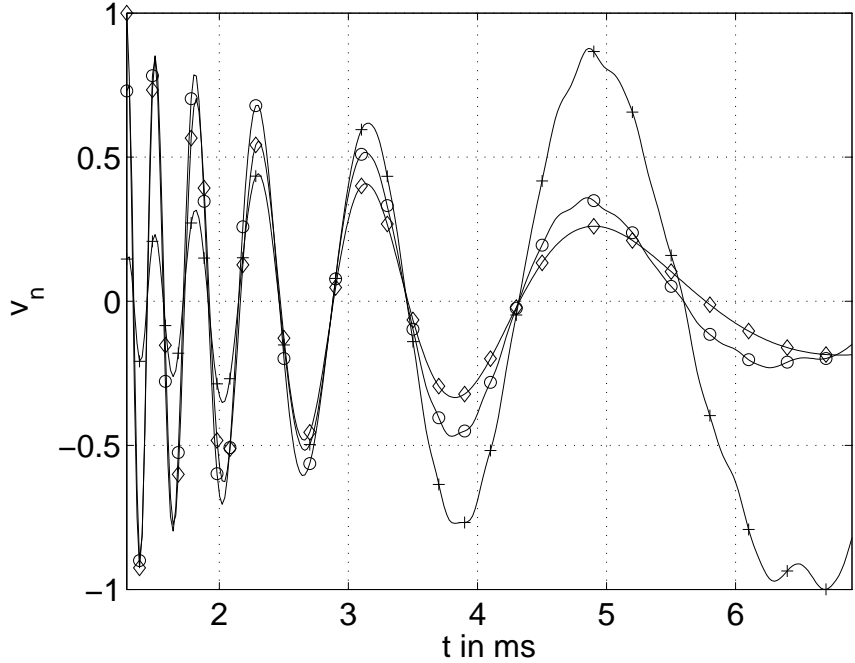}
\caption{Impulse response of a plate, normalised velocity $v(r=0.5m,t)/|v(r=0.5m,t)|_{max}$: measured (plus), theoretical (diamond) with equation (\ref{schnelleplatte}), measured and corrected with the power spectrum of the impulse (circle)}
\label{velo50run2}
\end{figure}

The dimensions of the plate are a thickness of $d=2mm$, a length of $l=2.05m$ and a width of $b=1.52m$. The material parameters provided by the manufacturer are elastic modulus $E=3.3GPa$,  $\rho=1190 kg/m^3$ and a Poisson's ratio $\nu=0.37$. \\
The whole plate is suspended in a frame. The velocity is measured with a laser-vibrometer at positions at least $0.5m$ from the frame. The excitation point is at a distance of $0.5m$ from the response position. \\ 
The velocity in figure \ref{velo50run2} is obtained in almost the same manner as in the beam experiment. The theoretical curve is calculated with equation (\ref{schnelleplatte}) and a value of $d_i=0.0544$ in a frequency range from $f_{min}=180 Hz$ to $f_{max}=5.2 kHz$. 

\section{Time reversal}
\begin{figure}
\includegraphics[width=0.9 \textwidth]{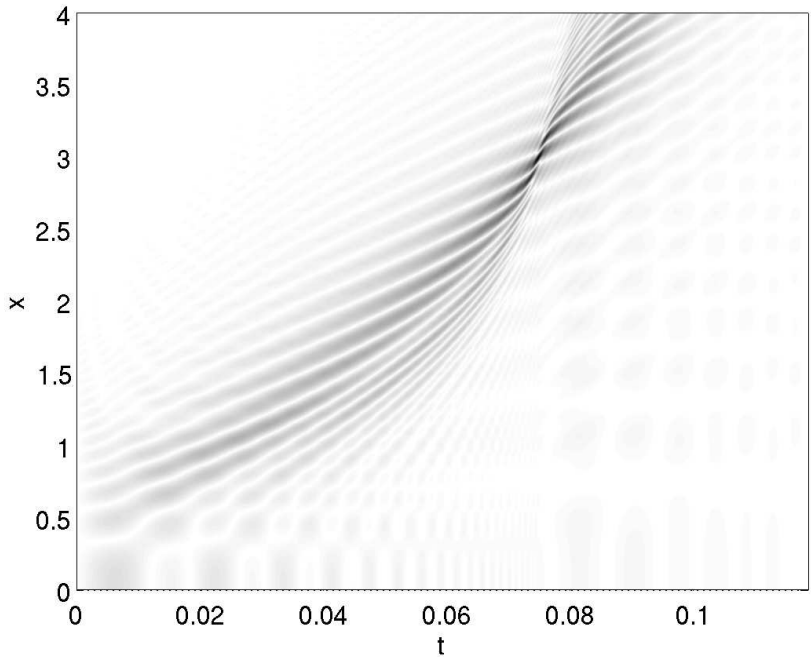}
\caption{Velocity of a beam excited with a time reversed pulse defined in equation (\ref{F_a_reverse}), $\zeta=1$, $x=3$. The darkness is given by $|v|$.}
\label{reziImpulse}
\end{figure}

To generate a defined pulse at a certain distance from the point excitation one may develop the time reversal from equation (\ref{vdirac}). As an example, a Dirac like pulse is requested at a position $x$ on a structure. By considering the force 
\begin{equation}   
F_a(t) = \left\{ \begin{array}{ll}
 \frac{F_0}{\sqrt{t_{max} - t}} \cos \frac{ x^2}{4 \zeta (t_{max} - t)}, & \mbox{ for } t_{min} < t < t_{max} \\
 0 & \mbox{ otherwise }. 
\end{array} \right. 
\label{F_a_reverse}
\end{equation}
a pulse resembling a $\delta$-function can be realized. The generated pulse will not be a perfect $\delta$-function that consists of the whole frequency spectrum, but will be a band-filtered version. \\
With equation (\ref{bochnerfreq}) the frequency range of equation (\ref{F_a_reverse}) is given by
\begin{equation}
f_{max/min} = \frac{x^2}{8 \zeta \pi t_{min/max}^2}.
\label{freqsin}
\end{equation}
In a numerical experiment the velocity is calculated with equation (\ref{vfaltung}) and shown in figure \ref{reziImpulse}. 

\subsection{Experimental Results}
\begin{figure}[ht]
\includegraphics[width=0.9 \textwidth]{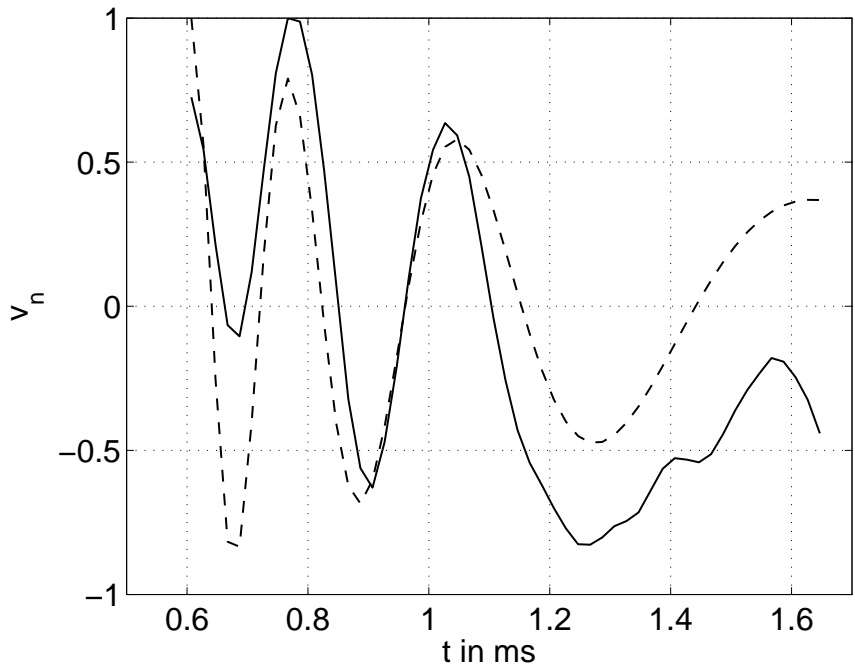}
\caption{Normalised velocity of a beam excited with a time reversed pulse defined in equation (\ref{F_a_reverse}) $v(x=1.13m,t)/|v(x=1.13m,t)|_{max}$ : measured (dashed), theoretical (solid) with equation (\ref{vdirac}) for a distance of $x_v = 0.73 m$}
\label{rezivn}
\end{figure}

The studied beam is the same like in section \ref{expbeam}, but now the beam is clamped at one end and hanging on a twine on the other. The measured velocity is analysed in the same manner as in the previous experiments. \\
 The generation of the desired force excitation is done with a magnetic force transducer. The dispersion factor for the reversed pulse (\ref{F_a_reverse}) is chosen to $d_i=0.0054$ with the given material properties of the beam the location of the pulse is at $x_{imp} = 0.4m$. The velocity is measured with a laser vibrometer in a distance of $x_{meas} = 1.13m$. It is assumed that from the receiver position a virtual source at a distance of $x_v = 0.73 m$ is seen. \\
The measured velocity is analysed with the method that is discussed subsequent publication and a dispersion factor of $d_i=0.0181$ is extracted within a frequency range between $f_{max} = 7.8 kHz$ and $f_{min} = 1.0 kHz$. The value of $d_i=0.0181$ corresponds with the distance of the virtual source of $x_v = 0.73 m$. The normalised velocity is plotted in figure \ref{rezivn}. The theorectical curve is calculated with the dispersion factor of $d_i=0.0181$. The results are not that clear as in the previous experiments. It is assumed that this is due to the comparable weak magnetic force transducer. 

\section{Concluding Remarks}
The main focus of this study is on the derivation of the impulse response
(\ref{vdirac}) and its application. It is a simple and instructive example of
dispersive wave propagation. \\
The applications of the impulse response function for time reversal is 
shown in theory and experiment. The attempt differs since an
analytical function is used and not the measurement impulse response. \\ 
Other applications in the fields of time frequency analysis and SEA are
indicated. This applications are rather roughly sketched and are subjects of
ongoing research. \\
Another natural extension is to study the conversion from transient to stationary
motion.    
\bibliographystyle{elsart-num}
\bibliography{lite}

\end{document}